\documentclass[aps,prb,floats,twocolumn]{revtex4}
\usepackage{epsfig}
\renewcommand{\vec}{\mathbf}

\begin{document}
\title{Correlation and surface effects in Vanadium oxides} 
\author{S.~Schwieger, M.~Potthoff and W.~Nolting}
\affiliation{Humboldt-Universi{\"a}t zu Berlin, Institut f{\"u}r Physik,
  Invalidenstr. 110, 10115 Berlin}
\begin{abstract}
Recent photoemission experiments have shown strong surface
modifications in the spectra from vanadium oxides as ${\rm (V,Cr)_2O_3}$ or ${\rm
  (Sr,Ca)VO_3}$. The effective mass is enhanced at the
surface and the coherent part of the surface spectrum is narrowed as
compared to the bulk. The
quasiparticle weight is more sensitive at the surface than in the bulk against bandwidth variations. We
investigate these effects theoretically considering the
single-band Hubbard model for a film geometry. 
A simplified dynamical mean-field scheme is used to calculate the main
features of the interacting layer-dependent spectral function.
It turns out that the experimentally confirmed effects are inherent
properties of a system of strongly correlated electrons. The reduction
of the weight and the variance of the coherent part of the surface spectrum can be traced
back to the reduced surface coordination number. Surface correlation effects
can be strongly amplified by changes of the hopping integrals at the surface.

\end{abstract}
\maketitle
\section{introduction}
A large class of transition-metal oxides display a metal-insulator
transition (MIT) upon variation of pressure, temperature or chemical
doping. According to effective one-particle theories these materials are normal
metals. The observed transitions are caused by correlations. Typical
materials that show such a behavior are vanadium oxides. There are numerous experimental \cite{IFT98} and
theoretical \cite{Geb97,Mot90} studies which describe and try to explain the physics of
these materials. The electronic structure in the vicinity of
the transition has been revealed  by photoemission experiments\cite{IFT98}. They
show a certain internal structure of the 3d-derived bands on the
metallic side of the transition. 
Essentially, there are two peaks in the angle-integrated spectral
function $A(\omega)$: One is located at the Fermi energy and is usually
called "coherent part" of the spectrum. The other structure , called
"incoherent part", is located at higher binding energies. The metal-insulator
transition is then driven by a redistribution of spectral weight from the
coherent to the incoherent peak. At the critical point the weight of the
coherent peak vanishes and the system undergoes the transition to the insulating phase. 

While an effective one-particle theory as the local-density
approximation within density-functional theory (DFT-LDA) can reproduce the coherent
part of the spectrum apart from a renormalization factor, it
fails to describe the incoherent part, which is caused by strong
electron correlations. 
On the other hand, the
correlation-induced MIT is a classical subject of many-body theory. If there is no
long-range magnetic order in the metallic as well as in the insulating phase, the
transition is known as the Mott-Hubbard
transition\cite{Mot49}. For band models there are early
theories that describe the incoherent \cite{Hub64} or the coherent
\cite{BrR70} part of the spectrum. 
Nowadays, with the dynamical mean-field theory
(DMFT)\cite{MeV89,JaR92,GKK96}
there is a general approach at hand which includes and unifies both
aspects. The DMFT reproduces the coherent (usually called quasiparticle
resonance) as well as the incoherent (usually called lower Hubbard band, LHB)
part of the spectrum. There is a third feature, the upper Hubbard band
(UHB), which is not seen in photoemission experiments since it is located
well above the Fermi energy. The DMFT yields a MIT as described
above\cite{GKK96,BCV01}. For bulk systems LDA+DMFT calculations even can give
quantitative results.\cite{KPA01,HKE01}

Recently, however, a number of studies have shown that the photoemission
data have to be interpreted carefully\cite{KPA01,MMS98,MSR01,SFI02}. Especially, the surface
sensitivity of photoemission has been reconsidered. It has been shown
that a proper interpretation of the data cannot rely on the presumption
that the electronic
structure at the surface is almost the same as in the bulk.
In fact, by comparing results from measurements with
different surface sensitivities, it has been shown that the spectral function
can be strongly modified at the surface compared with the bulk. It has been the intention of these studies
to extract reliable bulk properties from the experimental data which can
be compared with the results of LDA+DMFT theory, for example. However, also the
layer dependence and surface modifications of the spectral function have
been measured. These are interesting on its own. A number of qualitative trends can be deduced safely:
\begin{itemize}
\item The weight of the coherent part of the spectral function is 
  reduced at the surface. 
\item The width or more accurately the variance of the coherent peak
  tends to be somewhat
  smaller at the surface. 
\item Changes of the bandwidth (and thus of the effective correlation
  $\frac{U}{t}$) affect the weight of the coherent part of the spectral
  function much more at the surface than in the bulk.
\end{itemize}
It is an interesting theoretical task to explain these trends and to reproduce
the respective features in the spectral function $A(\omega)$.

In this paper we consider a system in a film geometry which exhibits a
Mott-Hubbard MIT and investigate the layer dependence of the spectral
function. The interesting question is whether the experimental findings are
inherent properties of a film geometry of correlated electrons or
whether additional modifications at the surface are required, such as
surface relaxation, surface phase separation or surface
reconstruction. If the measured trends are generic they should be present in
any model system that shows a Mott-Hubbard transition and has a layer
geometry with surface(s). Thus, though the investigated materials
usually require the study
of multi-band models to account for the band degeneracy,  the surface
modifications and layer dependence of the spectral function can be
investigated qualitatively using a single-band Hubbard model in a
film geometry. The advantage is that a (semi-) analytical approach can
be used which
makes the physical mechanisms beyond the surface effects most
explicit.

Of course, such a simple (semi-) analytical theory has to retain the fundamental
physics of the Mott-Hubbard transition as described above. This rules out theories as static mean-field theory or
Hubbard-III approximation \cite{Hub64} since they do not give the
correct three-peak structure of the spectral function. The three-peak structure of
the spectral function or equivalently a two-peak structure of the
imaginary part of the self energy is crucial for a qualitatively correct
(mean-field)
description of the Mott-Hubbard transition. A recent study \cite{BBC02}
has shown that experimental data can nicely be fitted once
this condition is met. Here we will use the recently developed
"two-site" DMFT \cite{Pot01}. 
This is an approach which keeps the essence of the DMFT but simplifies
the mean-field equations by a physically motivated approximation to allow for
analytical calculations. The two-site DMFT is an extension of the linearized
DMFT  \cite{BuP00} which meets the minimal condition mentioned above,
i.e. leads to a three-peak structure of the spectral function with two
Hubbard bands (incoherent peaks) and a quasiparticle resonance (coherent
peak).

In the next section we  specify the model assumptions and develop and
discuss the two-site DMFT for a film geometry. In section III we
discuss the reduction of the weight and the variance of the coherent
spectral function. Section IV considers the topic of the enhanced
sensitivity of the quasiparticle weight at the surface against bandwidth modifications. 
 
\section{Model and Theory}
We investigate the single-band Hubbard model for a film geometry  
\begin{equation}
H=-\sum_{ij\alpha\beta\sigma} t_{ij}^{\alpha\beta}
      c_{i\alpha\sigma}^+c_{j\beta\sigma} + \sum_{i\alpha\sigma}
      \frac{U_\alpha}{2}n_{i\alpha\sigma}n_{i\alpha-\sigma}
\label{Hamiltonian}
\end{equation}
Here $\alpha,\beta=1,\ldots,d$ label the different layers parallel to
the film surface(s). $d$ is the film thickness. The subscripts $i$ and
$j$ refer to the sites within a layer and run from $1$ to $N_\|$, where $N_\|$ is the
number of sites per layer ($N_\|\rightarrow\infty$).
In the following the on-site energies  and the
interaction strengths are taken to be layer independent
$t_{ii}^{\alpha\alpha}=t_0$ and $U_\alpha=U$.
$t_0=0$ defines the energy zero.
 The
hopping integrals $t_{ij}^{\alpha\beta}$ are assumed to be non-zero between nearest
neighbors only and is taken to be layer independent, too. The
nearest-neighbor hopping defines the energy unit
$t_{ij}^{\alpha\beta}=t=1$.
Only at the surface the
intra-layer hopping $t_{ij}^{11}=t_{11}$ or the hopping between the surface and
the subsurface layer $t_{ij}^{12}=t_{12}$ may be modified 
($t_{11},t_{12}\neq t$) to simulate relaxation processes.

 Exploiting the two-dimensional
translational symmetry, the layer-dependent Green function reads
\begin{equation}
G_{\alpha\beta,\sigma}(\vec{k}_\|,\omega) = \left[
\omega+\mu-\tilde{t}({\vec{k}_\|})-\tilde{\Sigma}_\sigma(\vec{k}_\|,\omega)\right]_{\alpha\beta}^{-1}
\label{Green}
\end{equation}
$\vec{k}_\|$ is a wave vector of the two-dimensional Brillouin zone. The
$d\times d$
matrix $\tilde{t}({\vec{k}_\|})$ is the Fourier-transformed hopping matrix
\begin{equation}
t_{\alpha\beta}(\vec{k}_\|)=\frac{1}{N_{\|}}\sum_{ij}t_{ij}^{\alpha\beta}e^{-i\vec{k}_\|(\vec{R}_i-\vec{R}_j)}
\label{Fourier}
\end{equation}

Within the DMFT the self-energy is local. The matrix $\tilde{\Sigma}(\vec{k}_\|,\omega)$ is
diagonal and $\vec{k}_\|$ independent. In the following the spin index
$\sigma=\uparrow,\downarrow$ is
dropped since we are solely interested in the paramagnetic phase.
The crucial point of the DMFT is the mapping of the lattice (Hubbard) model onto
an appropriate single-impurity (Anderson) model (SIAM). The latter is defined in
such a way that the impurity Green function and the self-energy of the
impurity model are equal to the on-site Green function and the
self-energy of the lattice model, respectively \cite{GKK96}. The DMFT is
exact in the limit of infinite spatial dimensions $D$ but can be applied as
a proper mean-field theory to finite-dimensional systems, too.

The formulation of the DMFT for a film geometry is straightforward\cite{PoN99a}:
The mapping has to be done
for each layer. Consequently in our case $d$ different
impurity models $H_{\rm imp}^{(\alpha)}$, one for each layer $\alpha$,
have to be defined.
The impurity models can be solved independently for $\alpha=1,
\ldots d$ but are coupled indirectly by a set of $d$ self-consistency
relations \cite{PoN99a}. As the usual DMFT for a bulk system, the
DMFT for a film geometry becomes exact for $D\rightarrow\infty$.

Within the
two-site DMFT \cite{Pot01} the mapping procedure is strongly
simplified. The respective single-impurity Anderson
models are replaced by  models that consist of one correlated impurity
site and one bath site only $H_{imp}^{(\alpha)}\rightarrow H_{\rm
  2-site}^{(\alpha)}$.
  This allows for an exact solution of
the impurity model. There are four parameters in the two-site model, two
of them are already fixed: the energy level of the correlated site
$\epsilon_d^{(\alpha)}\stackrel{!}{=}t_0=0$ and the interaction strength
$U_\alpha=U$. The remaining two
parameters, the hybridization between the two orbitals (sites)
$V_\alpha$ and the one-particle energy of the bath site $\epsilon_c^{(\alpha)}$, are chosen such that the original self-consistency conditions
are fulfilled in an integral way for the full spectral function and
especially for its coherent part. A
calculation completely analogous to the bulk case (see Ref. 17)
yields for hybridization $V_{\alpha}$:
\begin{equation}
{V_{\alpha}}^2 = \sum_{j\beta}
\left(t^{\alpha\beta}_{ij}\right)^2 z_{\beta}, \quad
(j\beta)\neq (i\alpha).
\label{V-mapping}
\end{equation}
Here the definition 
\begin{equation}
z_{\alpha}=\left(
  1-\Big[\frac{d\Sigma_\alpha(\omega)}{d\omega}\Big]_{\omega=0}\right)^{-1}
\label{z}
\end{equation}
is used. For metals $z_{\alpha}$ is the quasiparticle weight. Its inverse $z_{\alpha}^{-1}$
is the quasiparticle mass-enhancement factor.
At the critical interaction for the metal-insulator transition there is
a divergence of the effective mass:
$z_\alpha({U_c})\stackrel{!}{=}0$.
 
Finally, the one-particle energy of the bath site $\epsilon_{c}^{(\alpha)}$ is obtained from the second
self-consistency condition:
\begin{equation}
n^d_\alpha = n_{\alpha}
\label{eps-mapping}
\end{equation}
$n^d_\alpha$ and $n_{\alpha}$ are the particle densities for the
correlated site in the impurity model $H_{imp}^{(\alpha)}$ and for a site in the
$\alpha$-th layer of the lattice model, respectively.

Eqs. (\ref{V-mapping}) and (\ref{eps-mapping}) define the parameters of
the (two-site) impurity
models. The latter can easily be solved numerically or (for half
filling) analytically. The self-energy of the two-site model $\Sigma_{2-site}^{(\alpha)}(\omega)$ is
identified with the self-energy of the respective layer 
$\Sigma_{\alpha}(\omega)$. Now at once new quasiparticle
weights $z_\alpha$ are given by Eq. (\ref{z}). Via Eq.
(\ref{Green}) and the spectral theorem one obtains the new particle
density for each layer $n_\alpha$. Therewith, a new set of impurity models
can be defined. This circle has to be
iterated until self-consistency is reached. 

Almost analytic calculations
are possible for the symmetric case of the 
paramagnetic phase at half-filling. Eq. (\ref{eps-mapping}) is now
trivially fulfilled since $n^d_\alpha=n_\alpha=1$ due to
the manifest particle-hole symmetry, which requires $\epsilon_c^{(\alpha)}=\mu=\frac{U}{2}$. Furthermore, the
two-site problem can be solved analytically, which gives the self-energy
\begin{equation}
\Sigma_{\alpha}(\omega)=\frac{1}{2}U+\frac{\frac{1}{8}U^2}{\omega-3V_{\alpha}}+\frac{\frac{1}{8}U_\alpha^2}{\omega+3V_{\alpha}}
\label{sigma}
\end{equation} 
Thus Eq. (\ref{z}) can be evaluated
\begin{equation}
z_\alpha = \frac{36 V_{\alpha}^2}{36
  V_{\alpha}^2 + U^2}
\label{z-sym}
\end{equation}
Introducing the coordination number within a layer $q$ and between
two layers $p$ (bulk coordination number: q+2p), the mapping condition (\ref{V-mapping}) can be written
more explicitely
\begin{equation}
{V_{\alpha}}^2=qt_{\alpha\alpha}^2z_\alpha+pt_{\alpha\alpha+1}^2z_{\alpha+1}+pt_{\alpha\alpha-1}^2z_{\alpha-1}
\label{V-mapping1}
\end{equation}
For a bulk-system, where all quantities are layer
independent, the last two equations give a single non-trivial solution
and one recovers the Brinkmann-Rice \cite{BrR70} result
\begin{equation}
z_b(U)=1-\frac{U^2}{U_c^2}\quad .
\label{zbulk}
\end{equation}
The critical interaction, however, is different and given by the
linearized DMFT \cite{BuP00} value
\begin{equation}
U_c=6t\sqrt{2p+q}\quad.
\label{U_c}
\end{equation} 

For a film geometry, Eqs. (\ref{z-sym}) and (\ref{V-mapping1})
constitute a set of nonlinear algebraic equations which can easily be
solved by numerical means - even for large
systems. Then the self-energy
$\tilde{\Sigma}(\omega)$ (Eq. (\ref{sigma})) and the Green function
$\tilde{G}(k_\|,\omega)$
(Eq. (\ref{Green})) can be calculated and finally also the
spectral function
\begin{equation}
A_{\alpha}(\omega)=-\frac{1}{N_\|}\frac{1}{\pi}Im\sum_{\vec{k}_\|}G_{\alpha\alpha}({\vec{k}_\|},\omega)
\label{A}
\end{equation}

To make contact with the experimental findings discussed above, we have
to determine
the weight and the variance of the coherent part of the
spectrum. These quantities can be obtained analytically: The coherent
peak is described by the "coherent Green function" which is obtained
by expanding the self-energy for small frequencies 
\begin{equation}
\Sigma_\alpha(\omega)=
{a}_\alpha+{b}_\alpha\omega+{\cal O}(\omega^2)
\label{sigcoh}
\end{equation}
and inserting this expression into Eq. (\ref{Green}). For the self-energy
(\ref{sigma}) the parameters $a_\alpha$ and $b_\alpha$ are given by
\begin{eqnarray}
a_\alpha&=&\frac{U_\alpha}{2}=\mu\nonumber\\
b_\alpha&=&-\frac{U_\alpha^2}{36{V_{\alpha}}^2}=1-z_\alpha^{-1}
\label{ab}
\end{eqnarray}
This yields for the coherent Green function
\begin{equation}
{G}^{(coh)}_{ij\alpha\beta}(\omega)=\left[\omega-\tilde{t}-\tilde{b}\omega\right]_{ij\alpha\beta}^{-1}
\label{Greencoh}
\end{equation}
where $\tilde{t}$ is the real-space hopping matrix.
The coherent spectral function is given by the imaginary part of the
on-site element of the coherent Green function matrix:
\begin{equation}
A_\alpha^{(coh)}(\omega)=-\frac{1}{\pi}Im\, G_{ii\alpha
  \alpha}^{(coh)}(\omega+i0^+)
\label{Acoh}
\end{equation}
Its weight $w_\alpha$ and its variance $\Delta_\alpha^2$ are determined
by the spectral moments
$M^{(n)}_\alpha=\int_{-\infty}^\infty
d\omega\,\omega^n\,A_\alpha^{(coh)}(\omega)$:
\begin{eqnarray}
 w_\alpha&=&M_\alpha^{(0)}\nonumber\\
\Delta^2_\alpha&=&\left(M_\alpha^{(2)}-{M_\alpha^{(1)}}^2\right){M_\alpha^{(0)}}^{-1}
\label{weight}
\end{eqnarray}
To obtain the spectral moments, we perform a high-frequency expansion of
the coherent Green function (see Ref. 20)
\begin{eqnarray}
G_{ii\alpha
  \alpha}^{(coh)}(\omega)&=&\frac{M_\alpha^{(0)}}{\omega}+\frac{M_\alpha^{(1)}}{\omega^2}+\frac{M_\alpha^{(2)}}{\omega^3}\nonumber\\
{}&=&\frac{z_\alpha}{\omega}+\frac{z_\alpha^2\sum_{\beta
    j}^{(\beta j)\neq (\alpha
    i)}{t_{ij}^{\alpha\beta}}^2z_{\beta}}{\omega^3}
\label{expansion}
\end{eqnarray}
This yields
\begin{eqnarray}
w_\alpha&=&z_a\nonumber\\
\Delta^2_\alpha&=&z_\alpha\left(qt_{\alpha\alpha}^2z_\alpha+pt_{\alpha\alpha+1}^2z_{\alpha+1}+pt_{\alpha\alpha-1}^2z_{\alpha-1}\right)
\label{weight1}
\end{eqnarray}
With these results at hand we can analyze the surface effects
of the coherent part of the spectrum. 
Eqs. (\ref{z-sym}) and (\ref{V-mapping1}) turn out to be sufficient to discuss most
surface effects. Additionally, we can calculate the full spectral
function using Eq. (\ref{A}).

\section{Surface weight and variance}

\begin{figure}
\epsfig{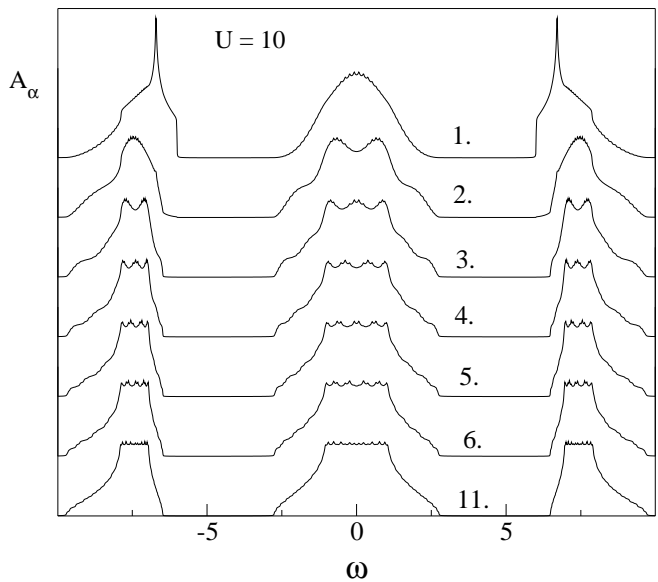}
\caption{Layer-dependent spectral function $A_\alpha(\omega)$ of a
  21 layer sc(100) film for uniform hopping $t=1$.}  
\label{Ueq10}
\end{figure}
Recall that the experiments show three characteristic surface
modifications of the spectral density: a reduced weight and a tendency
to a reduced variance of
the coherent part of the spectrum and an enhanced sensitivity against
bandwidth modifications at the surface.
Let us start with the first two phenomena, which have been observed for ${\rm (La,Ca)VO_3}$ (Ref. 13 ), for ${\rm (Ca,Sr)VO_3}$ (Ref. 14
and Ref. 15) and for
${\rm (V,Cr)_2O_3}$ (Ref. 11). We will discuss the spectra from ${\rm
  CaVO_3}$ and ${\rm SrVO_3}$
  (Ref. 14, 15) as representative examples. In the spectra
  of both materials there are two peaks, the coherent peak near the
  Fermi energy and the incoherent peak around 1.5 eV below $E_F$. This
  general structure is found for both, a surface-sensitive measurement
  as
  well as for a measurement with a weak surface sensitivity. The weight of the coherent peak is considerably
  reduced at the surface and its variance is somewhat smaller. The question we like to address is the
  following: Is the surface weight and variance reduction 
solely
caused by the reduced coordination number at the surface or are these
effects rather caused by modifications of surface parameters (which may
be due to e.g. surface relaxation or reconstruction).   
To this end we
  compare results for a Hubbard film with uniform hopping $t$ with those
  obtained from calculations for $t_{11},t_{12}\neq t$ at the film surface. 

{\bf Uniform parameters.}
We start the discussion with the case of uniform
  parameters.
Fig. \ref{Ueq10} shows the layer-dependent spectral function
$A_\alpha(\omega)$ of a 21 layer simple-cubic (100) (sc(100)) film for
$U=10$ and half-filling. Layer
"1" denotes the surface layer. The central layer ("11") simulates bulk
properties rather well. Each spectrum consists of a coherent peak around the Fermi
energy and incoherent peaks at $\omega\approx\pm 8$. In the
photoemission experiments only the occupied part of the spectrum is seen
which leads to the observed two-peak structure. Compared with the
experiments the different peaks are separated more clearly in Fig. \ref{Ueq10}. This is due
to the fact that
damping effects due to a finite imaginary part of the self energy are
neglected completely in the two-site DMFT.
\begin{figure}
\epsfig{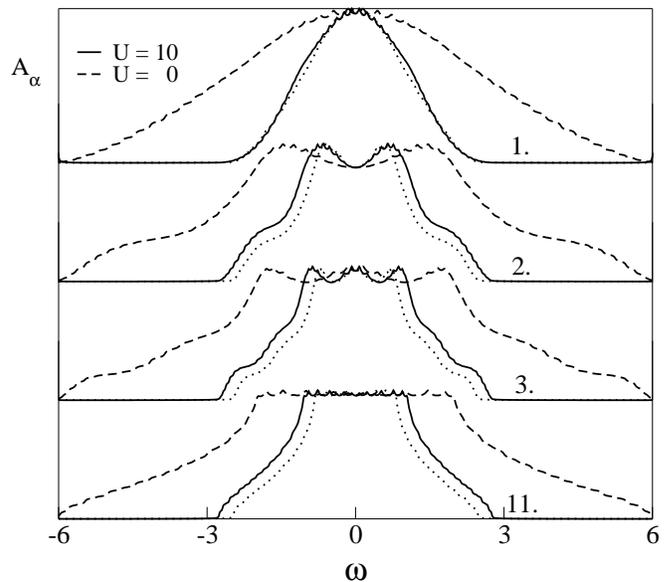}
\caption{Coherent part of the layer-dependent spectral function from Fig. \ref{Ueq10} (solid line),
  the free spectral function ($U=0$ dashed line) and the
  free spectral function narrowed by a constant factor $z=0.46$ (dotted line).  }
\label{U10U0}
\end{figure}
Damping effects are expected to be less important for the coherent peak
shown in Fig. \ref{U10U0}. As can be seen in the figure and as is also noticed in Ref. 15, the shape of the coherent
spectrum is well described by the uncorrelated ($U=0$) spectral function, apart
from a correlation-induced scaling factor. This factor $z$ can be
identified with the quasiparticle weight $z$ from Eq. (\ref{z}).
In
Fig. \ref{U10U0} comparison is made with the free spectral function narrowed by a constant
factor $z=0.46$ for all layers (dotted lines). While this factor works
well for the surface it is too small for the bulk. Hence the scaling
factor and therefore the quasiparticle weight is reduced at the
surface. This is the same trend as seen in the experiments.

However, the reduction is much weaker as in the experiments. A possible
reason is that the sc(100) surface is rather closed, i.e. the ratio
between the
surface and the bulk coordination number 
$\frac{p+q}{2p+q}=\frac{5}{6}$ is near unity.
\begin{figure}
\epsfig{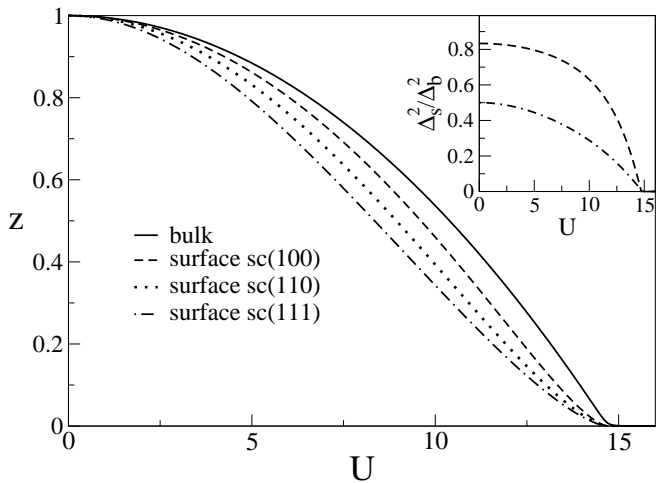}
\caption{$z$ as a function of $U$ in the bulk and at different surfaces
  of a simple cubic lattice for uniform parameters. Inset: the ratio of
  the surface and the bulk variance of the coherent peak as a function
  of $U$.}
\label{teqconst}
\end{figure}
In Fig. \ref{teqconst} we compare the quasiparticle weight $z$ of the
sc(100) surface with the ones of more open surfaces (sc(110) and sc(111)). Indeed
the surface quasiparticle weight is more reduced if the surface is less
closed. 

In the inset the ratio between the surface and the bulk variance is shown. The
surface variance is reduced for all interactions $U$ which is in
agreement with experiment. Again the reduction is more pronounced for
the more open sc(111) surface (dot-dashed line). We can distinguish
between a direct effect that is already present at $U=0$ and is given by the
ratio between the surface and the bulk coordination number (see
Eq. (\ref{weight1})) and an additional indirect
effect for finite interactions $U$
which is due to correlations.

The results discussed so far are not at all specific to the considered low-index
surfaces of a simple-cubic model structure. On the contrary, one can show
analytically that (within the two-site DMFT) the reduction of the weight
and the variance is inherent to any Mott-Hubbard system,
irrespective of the special geometry. We consider a
semi-infinite system ($d\rightarrow\infty$) with uniform parameters. 
We will show
that the weight at the surface $z_1$ is smaller than the weight
in the bulk $z_b$ ($0<z_1<z_b<1$ for $0<U<U_c$). 
From Eqs. (\ref{z-sym}) and (\ref{V-mapping1}) one readily derives the
recursion:
\begin{equation}
z_{\alpha+1}=\frac{z_\alpha}{1-z_\alpha}\frac{{U^\prime}^2}{p}-\frac{q}{p}z_\alpha-z_{\alpha-1},
\label{recursion}
\end{equation}
where $U^\prime=\frac{U}{6}$ is defined for convenience.
Now we find for the surface ($\alpha=1,z_0\equiv 0$)
\begin{equation}
z_1=1-\frac{{U^\prime}^2}{q+\frac{z_2}{z_1}p}
\label{z1}
\end{equation}
On the other hand, from (\ref{zbulk}) and (\ref{U_c}) we have for the bulk
\begin{equation}
z_b=1-\frac{{U^\prime}^2}{q+2p}
\label{zb}
\end{equation}
\begin{figure}
\epsfig{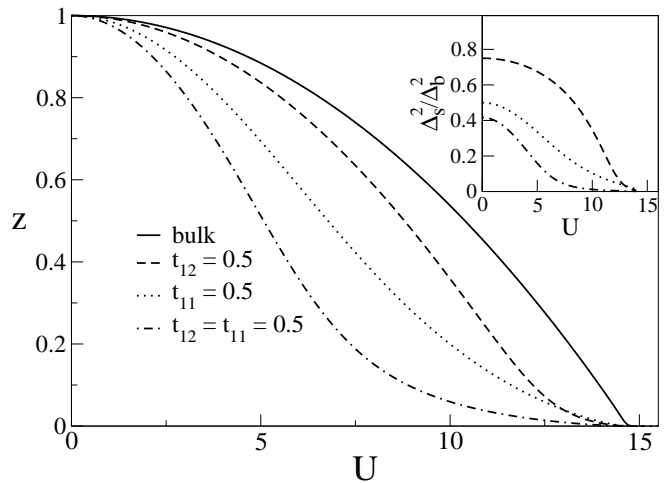}
\caption{$z_1$ and $z_{11}\approx z_b$ as functions of $U$ for a sc(100)
  film ($d=21$) and reduced hopping parameters at
  the surface. Inset: The ratio of surface and bulk variance of the
  coherent peak as
  a function of $U$.}
\label{zvonUtred}
\end{figure}

Now suppose for the moment that for a given $U$ ($0<U<U_c$)
\begin{equation}
z_1\geq z_b 
\label{vorr.}
\end{equation}
We will show that this leads to a contradiction.
It follows from (\ref{z1}) and (\ref{zb}) that the assumption (\ref{vorr.}) requires 
\begin{equation}
z_2>z_1\geq z_b\quad . 
\label{ianfang}
\end{equation}
Now
assume that the relation 
\begin{equation}
z_\alpha>z_{\alpha-1}\geq z_b
\label{vorr1}
\end{equation}
is
valid for arbitrary (fixed) $\alpha$. Then $\frac{1-z_b}{1-z_\alpha}>1$
holds. We immediately get
\begin{equation}
\frac{1-z_b}{1-z_\alpha}+\left(\frac{1-z_b}{1-z_\alpha}-\frac{z_{\alpha-1}}{z_a}\right)+\frac{q}{p}\left(\frac{1-z_b}{1-z_\alpha}-1\right)>1
\end{equation}
since the last two addends are positive. After some algebra one arrives
at
\begin{equation}
\frac{(1-z_b)(2p+q)}{(1-z_\alpha)p}-\frac{q}{p}-\frac{z_{\alpha-1}}{z_\alpha}>1
\end{equation}
Using (\ref{zb}) and multiplying the inequality with $z_\alpha$ gives
\begin{equation}
\frac{z_\alpha}{1-z_\alpha}\frac{{U^\prime}^2}{p}-\frac{q}{p}z_\alpha-{z_{\alpha-1}}>z_\alpha
\end{equation}
The left hand side of this expression is just $z_{\alpha+1}$ as seen
from Eq. (\ref{recursion}). Hence 
\begin{equation}
z_{\alpha+1}>z_\alpha\geq z_b
\label{result1}
\end{equation}
It is shown that (\ref{result1}) directly follows from (\ref{vorr1}) and
(\ref{vorr1}) holds for $\alpha=2$ (\ref{ianfang}).
Consequently (\ref{result1}) holds for all $\alpha$.

On the other hand $z_\alpha$ has to converge against the
bulk value $z_b$ for $\alpha\rightarrow\infty$
\begin{equation}
z_\alpha\stackrel{\alpha\rightarrow\infty}{\longrightarrow}z_b
\label{andererseits}
\end{equation}
The last two statements (\ref{result1}) and (\ref{andererseits}) are contradictory. Hence 
our assumption
(\ref{vorr.}), i.e. that $z_1$ equals or is greater than $z_b$, cannot
be valid. Thus it is shown that the quasiparticle
weight is reduced at the surface 
\begin{equation}
z_1<z_b \quad {\rm for} \quad 0<U<U_c\,,
\end{equation}
for uniform parameters, irrespective
of the geometry.

A short and vivid explanation of this
finding is based on the reduced coordination number at the
surface. Let us compare Eq. (\ref{V-mapping1}) for the surface and for
the bulk.
\begin{eqnarray}
{V_{1}}^2 &=& z_1q+z_2p\quad,\nonumber\\
{V_{b}}^2 &=& z_bq+z_bp+z_bp
\label{handwave1}
\end{eqnarray}
There is a (positive) addend missing in the surface expression. This
tends to reduce the surface parameter $V_{1}$ which causes a reduced
quasiparticle weight $z_1(U)$ in turn (Eq. (\ref{z-sym})).

All experiments discussed above show a reduced weight of the coherent
peak at the surface. Some of them may even be interpreted as showing an
insulating surface of a metallic bulk. However this is ruled out for uniform parameters. From
Eq. (\ref{recursion}) it follows immediately, that all layers are
insulating if the quasiparticle weight of the first layer vanishes. (Recall that $z_0=0$.) 

Similar considerations as for the weight apply for the variance.
From
Eq. (\ref{weight1} and (\ref{V-mapping1}) we have:
\begin{eqnarray}
\Delta_1^2&=&z_1V_{1}^2\nonumber\\
\Delta_b^2&=&z_bV_{b}^2
\label{variance}
\end{eqnarray}
Using $V_\alpha^2=\frac{z_\alpha}{1-z_\alpha}{U^\prime}^2$ (from
Eq. (\ref{z-sym})) and Eq. (\ref{zb}) we end up with
\begin{eqnarray}
\frac{\Delta_b^2}{2p+q}&=&z_b^2\nonumber\\
\frac{\Delta_1^2}{2p+q}&=&\frac{1-z_b}{1-z_1}z_1^2
\label{variance1}
\end{eqnarray}
Since we already know $0<z_1<z_b<1$ this immediately yields
\begin{eqnarray}
\Delta_1^2<\Delta_b^2\nonumber
\end{eqnarray}
for any surface geometry and for uniform but arbitrary parameters. 

Again, a closer
look at (\ref{z-sym}) and (\ref{V-mapping1}) gives a vivid
explanation of this finding: Starting from (\ref{weight1}), we get for the
surface and the bulk variance:
\begin{eqnarray}
\Delta_1^2&=&z_1(qz_1+pz_2)\quad,\nonumber\\
\Delta_b^2&=&z_b(qz_b+pz_b+pz_b)\nonumber\\
\label{handwave2}
\end{eqnarray}
Here one can see both effects that have been already discussed at
Fig. \ref{teqconst}.  The {\bf direct} effect is
just the missing addend in the surface expression (similar to (\ref{handwave1})).
This reduces the surface variance even for $U=0$. The {\bf indirect } effect, which is due to correlations, is caused by
the reduced surface weight $z_1(U)<z_b(U)$. Both effects add and result
in a reduced variance at the surface as seen in Fig. \ref{teqconst}.

{\bf Modified surface parameters.}
Even for the relatively open surface sc(111) the reduction of the quasiparticle
weight at the surface is smaller as compared to the reduction found in the experiments. We therefore
want to investigate whether a modified hopping at the surface (caused
e.g. by a surface relaxation) can further reduce the
spectral weight and the variance of the coherent peak.  
Indeed Fig. \ref{zvonUtred} shows a considerable influence of the
surface hopping on the quasiparticle weight. If both, the hopping within the
surface layer $t_{11}$ and the hopping between the surface and the
subsurface layer $t_{12}$ are reduced (dot-dashed line) the weight of the coherent peak is
almost negligible for a wide range of interactions strengths.
\begin{figure}
\epsfig{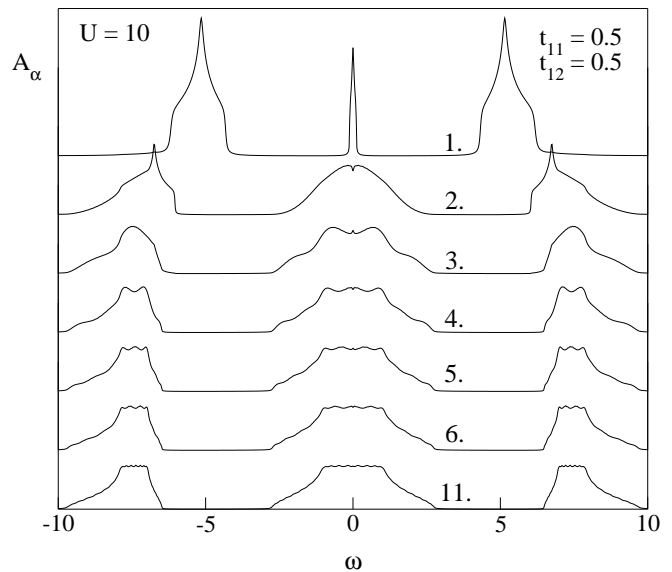}
\caption{As Fig. \ref{Ueq10} but for reduced hopping parameters at the surface.}
\label{t1eq05}
\end{figure}
This applies e.g. for $U=10$ which is still far away from the bulk
metal-insulator transition. Nevertheless, looking at the layer dependent
spectral function (Fig. \ref{t1eq05}), there are strong correlation
effects at the surface. Now there is only a very narrow
peak at the Fermi energy for $\alpha=1$. This fits well to the experiments
cited above. The inset of Fig. \ref{zvonUtred} shows analogous
trends for the variance.

Summing up, one can state that both experimentally established trends,
i.e. the reduced weight and the reduced variance of the coherent surface peak, are
inherent features of a Mott-Hubbard system in a film geometry. Reduced
hopping integrals at the surface may amplify both effects considerably.

\section{Sensitivity against bandwidth variations}

Another surface effect is mentioned in Ref. 15 for ${\rm
  (Sr,Ca)VO_3}$. 
Again, measurements with strong as well as with 
weak surface sensitivity have been performed. It is found that the surface
weight $z_1$ is much smaller in ${\rm CaVO_3}$ than in ${\rm SrVO_3}$
while the quasiparticle weight in the bulk $z_b$ is almost the same. 
Both materials are expected to exhibit comparable interactions $U$.
The
difference is mainly the hopping integral $t$ 
which is smaller in ${\rm CaVO_3}$. In other words, 
the experiments show that
the slope of the function $z(t)$ - $\frac{dz}{dt}$ - is
positive and much higher at the surface compared to the bulk.
\begin{figure}
\epsfig{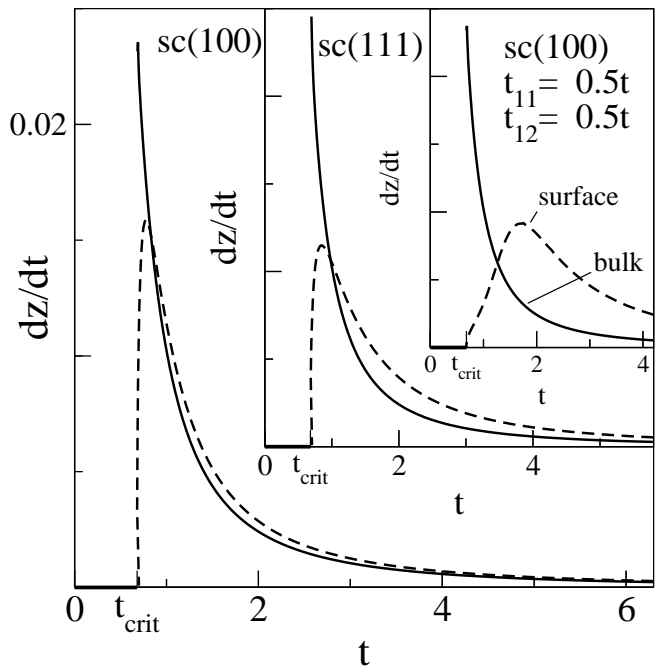}
\caption{The slope of the quasiparticle weight as a function of the
  hopping parameter $t$ for constant $U=10$.}
\label{dzdt}
\end{figure}
Fig. \ref{dzdt} shows this slope as a function of the hopping $t$
(and thus as a function of the bandwidth $W=12t$) at a
fixed interaction $U=10$. Below the critical hopping $t_{crit}=0.68$
the system is insulating. For the first case, a sc(100)
surface, we can qualitatively reproduce the experimental result for a wide parameter range
($t>0.85$). However the enhancement is very weak. As for the other
surface effects, the enhancement becomes stronger for more open surfaces
as well as for reduced hopping integrals at the surface. The
experimental finding $\frac{dz_1}{dt}>\frac{dz_b}{dt}$ is thus expected
for most parameters. However, the opposite scenario may be found as
well for smaller hopping integrals $t$.

\section{summary}
We have reproduced and explained a number of surface effects recently
detected by photoemission experiments from metallic Vanadium
oxides. Within a simplified but reasonable DMFT scheme, it was shown
analytically and numerically that the reduction of the quasiparticle
weight and the variance of the coherent part of the spectrum at the surface are inherent properties of a system of
correlated electrons. It was explicitely shown that these effects are caused by the reduced surface coordination
number. Consequently, the surface modifications are stronger for an open surface. These
effects can be amplified if the
hopping integrals at the surface are reduced.

The situation is not so clear for the enhanced sensitivity of the surface
quasiparticle weight $z_1$ against bandwidth variations. Though the
experimental trend $\frac{dz_1}{dt}>\frac{dz_b}{dt}$ is found for a wide
range of parameters, the opposite is true for parameters that are still
realistic. Furthermore we did not find any region in the parameter
space where $z_1$ changes considerably but $z_b$ is nearly insensitive
against bandwidth variations. This subject needs also more clarification
experimentally: In Ref. 14 for instance, where the same materials are
investigated as in Ref. 15, the bulk quasiparticle weight still seems to
change if one compares $SrVO_3$ and $CaVO_3$.

\section*{Acknowledgments}
This work is supported by the Deutsche Forschungsgemeinschaft within
the Sonderforschungsbereich 290.


\begin{thebibliography}{10}
\bibitem{IFT98}
for a review see:
M.~Imada, A.~Fujimori and Y.~Tokura, Rev.~Mod.~Phys. {\bf 70}, 1039 (1998) 
\bibitem{Geb97}
for a review see:
F.~Gebhard, {\it The Mott Metal-Insulator Transition}, (Springer, Berlin
1997)
\bibitem{Mot90}
N.~F.~Mott, {\it Metal-Insulator Transitions}, (Taylor and Francis,
London,1990)
\bibitem{Mot49}
N.~F.~Mott, Proc.~Roy.~Soc.London A {\bf 62}, 416 (1949)
\bibitem{Hub64}
J,~Hubbard, Proc.~Roy.~Soc.London A {\bf 281}, 401 (1964)
\bibitem{BrR70}
W.~F.~Brinkman and T.~M.~Rice, Phys.~Rev.~B {\bf 2}, 4302 (1970)
\bibitem{MeV89}
W.~Metzner and D.~Vollhardt, Phys.~Rev.~Lett. {\bf 62}, 324 (1989)
\bibitem{JaR92}
M.~Jarrell, Phys.~Rev.~Lett. {\bf 69}, 168 (1992)
\bibitem{GKK96}
A.~Georges, G.~Kotliar, W.~Krauth and M.~J.~Rozenberg,
Rev.~Mod.~Phys. {\bf 68}, 13 (1996)
\bibitem{BCV01}
R.~Bulla, T.~A.~Costi, and D.~Vollhardt,
Phys.~Rev.~B {\bf 64}, 045103 (2001)
\bibitem{KPA01}
Hyeong-Do~Kim, J.~H.~Park, J.~W.~Allen, A.~Sekiyama, A.~Yamasaki,
K.~Kadono, S.~ Suga, Y.~Saitoh, T.~Muro, P.~Metcalf, cond-mat/0108044
\bibitem{HKE01}
K.~Held, G.~Keller, V.~Eyert, D.~Vollhardt and V.~I.~Anisimov,
Phys.~Rev.~Lett. {\bf 86}, 5345 (2001)
\bibitem{MMS98}
K.~Maiti, P.~Mahadevan, and D.~D.~Sarma,
Phys.~Rev.~Lett. {\bf 80}, 2885-2888 (1998)
\bibitem{MSR01}
K.~Maiti, D.~D.~Sarma, M.~J.~Rozenberg, I.~H.~Inoue, H.~Makino, O.~Goto,
M.~Pedio and R.~Cimino,
Europhys.~Lett. {\bf 55}, 246 (2001) 
\bibitem{SFI02}
A.~Sekiyama, H.~Fujiwara, S.~Imada, H.~Eisaki, S.~I.~Uchida,
K.~Takegahara, H.~Harima, Y.~Saitoh, S.~Suga,
cond-mat/0206471
\bibitem{BBC02}
 K.~Byczuk, R.~Bulla, R.~Claessen, D.~Vollhardt,
cond-mat/0205149
\bibitem{Pot01}
M.~Potthoff, Phys.~Rev.~B {\bf 64}, 165114 (2001)
\bibitem{BuP00}
R.~Bulla, M.~Potthoff,
Eur.~Phys.~J.~B {\bf 13}, 257 (2000)
\bibitem{PoN99a}
M.~Potthoff and W.~Nolting, Phys.~Rev.~B {\bf 59}, 2549 (1999);
Eur.~Phys.~J.~B {\bf 8}, 555 (1999)

\bibitem{PHW98}
 M.~Potthoff, T.~Herrman, T.~Wegner, and W.~Nolting,
phys.~stat.~sol.~(b) {\bf 210}, 199 (1998)

\end{thebibliography}
\end{document}